# Van Der Waals Heteroepitaxy of GaSe and InSe, Quantum Wells and Superlattices

Marcel S. Claro†*[1,2], Juan P. Martínez-Pastor[3], Alejandro Molina-Sánchez[3], Khalil El Hajraoui[1], Justyna Grzonka[1], Hamid Pashaei Adl[3], David Fuertes Marrón[4], Paulo J. Ferreira[1], Alex Bondarchuk[1], Sascha Sadewasser[1,2]

[1] INL-International Iberian Nanotechnology Laboratory, Av. Mestre José Veiga s/n, 4715-330 Braga, Portugal.

[2] QuantaLab, 4715-330 Braga, Portugal

[3] ICMUV, Instituto de Ciencia de Materiales, Universidad de Valencia, P.O. Box 22085, 46071 Valencia, Spain

[4] Instituto de Energía Solar-ETSIT, Universidad Politécnica de Madrid, 28040, Madrid, Spain

*Corresponding author: marcel.santos@usc.es

*Bandgap engineering and quantum confinement in semiconductor heterostructures provide the means to fine-tune material response to electromagnetic fields and light in a wide range of the spectrum. Nonetheless, forming semiconductor heterostructures on lattice-mismatched substrates has been a challenge for several decades, leading to restrictions for device integration and the lack of efficient devices in important wavelength bands. Here, we show that the van der Waals epitaxy of two-dimensional (2D) GaSe and InSe heterostructures occur on substrates with substantially different lattice parameters, namely silicon and sapphire. The GaSe/InSe*

*heterostructures were applied in the growth of quantum wells and superlattices presenting photoluminescence and absorption related to interband transitions. Moreover, we demonstrate a self-powered photodetector based on this heterostructure on Si that works in the visible-NIR wavelength range. Fabricated at wafer-scale, these results pave the way for an easy integration of optoelectronics based on these layered 2D materials in current Si technology.*

# 1. Introduction

III-VI post-transition metal chalcogenides (PTMC), like GaSe, GaS, GaTe, InS, InSe, and InTe, crystallize in a hexagonal lattice of four monoatomic sheets known as tetralayers (TL), bonded by van der Waals (vdW) forces and surface lattice parameter around 4 Å (Figure 1a). Its bulk form has been known since the 1970s, and they have been used for some non-linear optical applications in the infrared[1,2], second harmonic generation[3] (SHG) or circular dichroism[4] in the meantime. Despite the unexplored properties as thin films, in the bulk form, the chemical, optical and electronic properties are already well studied. The bulk PTMC layered semiconductors present a direct bandgap of 0.6 - 2.6 eV, and present one of the highest electron mobility of layered materials[5], which would make them ideal for electronic and optoelectronic applications.

Some of the PTMCs can be found in several phases or polymorphs. Indium Selenide, for example, appears as InSe, α-$In_2Se_3$, β-$In_2Se_3$, $In_3Se_4$, etc.[6], which makes the synthesis of these materials challenging, but at the same time it opens the possibility of exploiting specific properties of each phase and the phase changes itself. Few memory devices based on phase transformation have been already reported[7–9]. β-$In_2Se_3$ and α-$In_2Se_3$ present robust ferroelectricity[10] at room temperature, which is a building block for energy harvesting/storage and neuromorphic devices, and still a rare property in the 2D material world.[11] The InSe, which we deal in this paper, apart from the already mentioned high-mobility, exhibited Quantum hall effect[12], good photoresponsivity, good mechanical strength, and flexiblility[13], exotic properties when strained[14], and low lattice conductivity (<2 Wm$^{-1}$K$^{-1}$)[15] together with high Seeback coefficient[6], important for thermoelectric (TE) materials, being the perfect definition of multifunctional material. Due to lattice and electronic band similarities, the equivalent properties are expected in GaSe and all PTMCs.

Yet, GaSe and InSe have several known polymorphs or polytypes with similar formation energy[16], appearing depending on the synthesis method. The most common polytypes are formed by non-centrosymmetric TL (D$_{3h}$), named as ε-(P$\bar{6}$m2), β-(P6$_3$/mmc), γ-(R3m), and δ-(P6$_3$mc), based on different stacking sequences between the adjacent layers along the [0001] orientation.[16–19]. Dissimilar polymorphs present small differences in band structure, electronic and optical properties particularities due to the presence or lack of special symmetries.

*Ab initio* studies and ARPES direct measurements have found Mexican-hat valence band (MHVB) dispersions in the valence band of many few-layer PTMCs[20–22]. The uncommon MHVB result in step-like and delta-like density of states (DOS), large transport velocity due to the flatness and steepness of the band, and electronic instabilities that lead to tunable superconductivity, ferroelasticity, and ferromagnetism. MHVB is present in the most efficient thermoelectric materials, e.g., Bi$_2$Se$_3$ and Bi$_2$Te$_3$.[23,24] It is anticipated that PTMC TE figures-of-merit can be improved when thinned to TL scale, due to the increment of the density of states (DOS) when the band structure turns into MHVB.[23,24] Such thin PTMCs in theory would present zT as high as 2.0[23]: as good as the reference Bi$_2$Te$_3$ alloys.

Most importantly, it was demonstrated that heterostructures from InSe and GaSe flakes exhibit optical transitions that densely cover the spectrum from violet to infrared[25], something that was rarely achieved with any other material system.

Decades ago, Esaki and Tsu[26] used the Molecular beam epitaxy (MBE) of two semiconductor compounds with distinct bandgaps, GaAs and AlAs, to create the first man-made artificial crystal using ultrathin layers of material. Layers with few nm forming quantum-wells (QW) and superlattices (SL) succeed on explore the effect of quantum tunneling and quantum confinement with unprecedented control since the carrier's potential barrier is determined by the bandgap difference, band offset, and the precisely chosen layer thickness. In the first few years, using the same materials, high-mobility transistors and optoelectronic devices (diode lasers and photodetectors) were invented, and the following years were devoted to expanding the family of the III-V semiconductors to obtain different bandgaps and band offsets for optoelectronics. Additionally to the technological applications on high-speed communication[27], energy[28], defence[29], health[30,31], or science[32], such control allowed the discovery of several novel quantum effects e.g. fractional[33] and spin[34] quantum hall effect, topological matter[35], etc. Nonetheless, the epitaxy technology has a major drawback: the lattice size mismatch between the different materials beyond a limit creates local strain that typically prevents the layer-by-layer growth for lattice mismatch higher than 1-3 %. In the first years, most of the epitaxial devices were based on GaAs or InP substrates, with lattice parameter of 5.65 and 5.85 Å respectively, and the III-V alloys (Ga, In, Al with P, As, Sb) with similar lattice sizes. Unfortunately, those heterostructures only reach efficiently the infrared and the red part of the visible spectrum, up to 1.5 eV. Widegap III-V technology over sapphire substrates was developed only in the 1990s[36] using Ga, In, and Al nitrides, when the first blue light-emitting diodes were invented. However, several important wavelengths and a whole region of the spectrum still lack a material with good properties for device fabrication. Even with new compound semiconductors developed for the green region, like the II-VI family (Zn, Mg, Cd with S, Se, and Te), we are still unable to achieve the required material quality, since doping and epitaxial growth of II-VI never prospered as expected. Figure 1b shows the lattice parameter and the bandgap of those materials where some observe the lack of direct bandgap materials in the green, orange and yellow window, and the complete void in the lattice parameter from 3.5 to 5.2 Å. As result, the only viable high-power alternatives in this color region[30,37] make use of frequency down or up-conversion in complex non-linear solid-state systems, which price, size, limitations in power, and wavelength coverage results unsuitable for wide exploitation.

Most recently, since the interest in 2D PTMC resurged due to its novel properties as 2D materials and its promises for optoelectronics, few groups report the growth of thin films by CVT[38,39], PVT[40,41], CVD[42,43], PLD[44], or Molecular beam epitaxy - MBE[45–48]. However, these efforts, with few exceptions, have been limited to small area characterization or few devices, rather than large-area, pretty much like as it is done with 2D material flakes, suggesting challenges in achieving wafer-scale uniformity. A fairly general trend in van der Waals epitaxy. The epitaxy by MBE of bulk PTMCs has been reported without constancy since the 1990s[49–53] and unfortunately, most of them reported only the homoepitaxy or the standard epitaxy on GaAs substrates, which results in rough interfaces and growth in other orientations than the [0001], and which cannot be directly applied for atomic defined interfaces as required for quantum wells and superlattices. Only recently, Sorokin demonstrate the growth of GaTe/GaSe[48,54] heterostructures which could be used for this goal, however, up to now, no one succeed to grow InSe/GaSe heterostructures or to demonstrated quantum confinement on PTMC large-area heterostructures.

Recently, we presented the epitaxial growth of high-quality and single-phase β-In$_2$Se$_3$ by MBE on a 2-inch wafer scale[55]. Thin layers down to two QLs were uniformly obtained and I fabricated photodetectors using photolithography and other standard semiconductor processing. In the same work, we demonstrated that changing substrate temperature and Se flux is possible to obtain different Indium Selenide phases, including the InSe phase. We also optimized the growth of GaSe and obtained high-quality thin layers on two completely different substrates: c-sapphire (4.7 Å) and Si (111) (3.85 Å)[56]. Using atomic-resolution scanning transmission electron microscopy (STEM), we investigated the formation of its multiple polytypes during the growth and correlated defects. It was observed that, far from the substrate, the GaSe and InSe relax and assume their original lattice parameter (3.75 and 4.0 Å, respectively). We also reported the first structural observation of a new polymorph ($R\bar{3}m$) (named as γ´-polymorph) characterized by a distinct atomic configuration with centrosymmetric TL ($D_{3d}$)[18,56,57].

In this work we have combined both GaSe and InSe epitaxy in heterostructures, growing short-period (~2 TL) InSe/GaSe superlattices and quantum-wells in both substrates (sapphire and Si), They presented photoluminescence which we tunned changing the QW widths demonstrating quantum confinement. It was observed that the InSe and GaSe lattices are commensurable in the superlattice, making both materials strained (with 3.88 Å) with the $D_{3d}$ polymorph dominant which probably presents enhanced properties in comparison with the non-centrosymmetric TL ($D_{3h}$)[18]. Finally, we were able to fabricate a simple photodetector based on this heterostructure at wafer scale, to demonstrate the applicability of the GaSe/InSe heteroepitaxy by MBE.

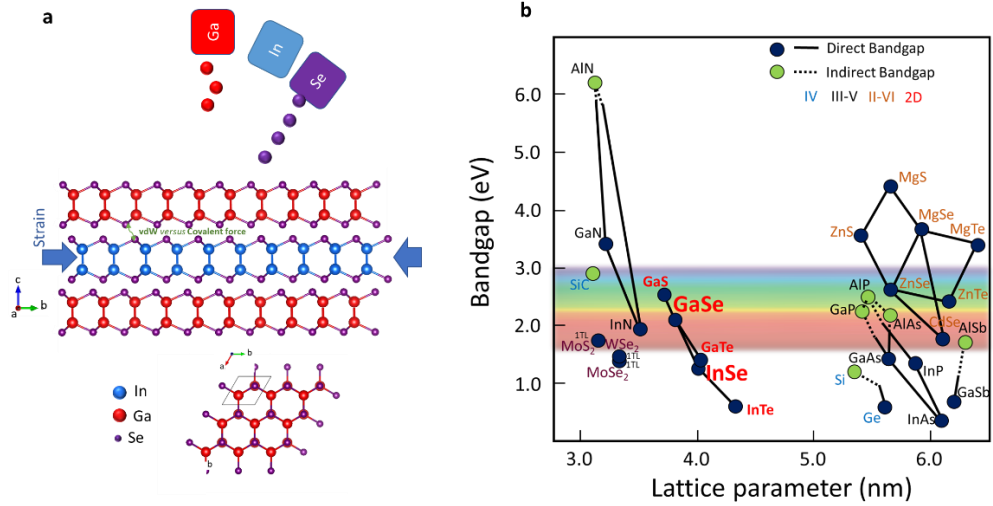

*Figure 1. a) Illustration of the growth by MBE of PTMC heterostructures detailing the crystal lattice b) Bandgap vs. lattice parameter diagram of the most common semiconductors and the PTMCs.*

## 2. Results

### 2.1 Large-area growth by MBE

Since gallium selenide ($Ga_xSe_y$) and indium selenide ($In_xSe_y$) exist in various solid phases, and form a complex phase diagram, a precise control on stoichiometry and growth conditions is required.[6] As demonstrated earlier by our group[55] and others[41,46], MBE provides the necessary means to control the III and VI material fluxes to obtain the desired pure phase, and in addition, to monitor the growth in-situ by reflection high energy electron diffraction (RHEED). For our samples, the MBE growth proceeds from elemental indium (6N) and gallium (6N) sources, evaporated from Knudsen cells, and selenium (5N), provided from a valved cracker cell wherein the flux is controlled by a valve with an adjustable aperture. The III-VI ratio is controlled by the Se cracker cell valve aperture, and the growth rate by the In and Ga Knudsen cell temperature. The lowest value of surface roughness was obtained within the window of substrate temperatures between 450 to 600 °C for (thick, > 60nm) multilayer growth. Starting the growth beyond this temperature range usually results in randomly oriented polycrystalline films, or their complete vaporization, as observed by in-situ reflection high energy electron diffraction (RHEED). Within this temperature window, three phases of $In_xSe_y$ were selectively obtained by varying the Se flux through the valve aperture (linear relation in the range used). InSe was obtained with the valve at 0.75 mm or less, γ-$In_2Se_3$ from 1 to 3 mm and β-$In_2Se_3$ from 2 to 5 mm.[55] GaSe was obtained with the valve at 1.2 mm or less, and $Ga_2Se_3$ beyond this value. In general, to obtain single-phase InSe (GaSe), In-rich (Ga-Rich) growth conditions are required. Despite the narrow growth window[48], the valve aperture is finely tuned for each temperature and material by monitoring the RHEED pattern. When the valve is found at the optimized value, the RHEED diffraction pattern (Figure 2a) is bright and streaky even after the growth of several layers, otherwise, the pattern fades in intensity and the streaky lines become diffuse. Very similar patterns are obtained on Si (111) substrates without surface treatments. However, highly-oriented GaSe can be obtained on Si (111) substrates taking advantage of the Ga-Se passivation layer (Figure 2b). Despite that, the crystallographic order was in general not sustained along with several TLs or within heterostructures, resulting in patterns as in Figure 2a. Advantageously, a single c-sapphire substrate could be used for calibration of the fluxes, as the substrate surface is immutable even after several growths and desorption cycles.

Initially, single heterostructures consisting of thick layers (~30 TL) of single phase InSe on top of thick (~60 TL) GaSe were grown on Silicon (111) and c-Sapphire substrates at the growth rate of approximately 1 TL/min. The Raman spectrum (Figure 2c) shows peaks at 133 cm$^{-1}$, 205 cm$^{-1}$, and 307 cm$^{-1}$, corresponding to the $A^1_1$, $E^1$, and $A^2_1$ vibration mode of GaSe, respectively[58], while the peaks located at 115cm$^{-1}$, 177 cm$^{-1}$, and 227 cm$^{-1}$ are attributed to the same modes for InSe[41]. No evidence of other phases, namely Ga$_2$Se$_3$ or In$_2$Se$_3$ is inferred. In addition, distinct and narrow peaks in X-ray diffraction (XRD) support the high quality of growth, including high-quality single phase InSe, which could not be obtained directly on bare Si and sapphire substrates, without a GaSe buffer layer[55] (Figure 2d). The surface topography obtained by atomic force microscopy (AFM) exhibits atomic terraces with triangular shape, as expected for these hexagonal crystals (Figure 2e), wherein the crystal grows laterally along the [$\bar{2}11$] or [$\bar{1}2\bar{1}$] directions. The small accumulation of Se in the surface which appears in the AFM as tall round particles will be discussed later.

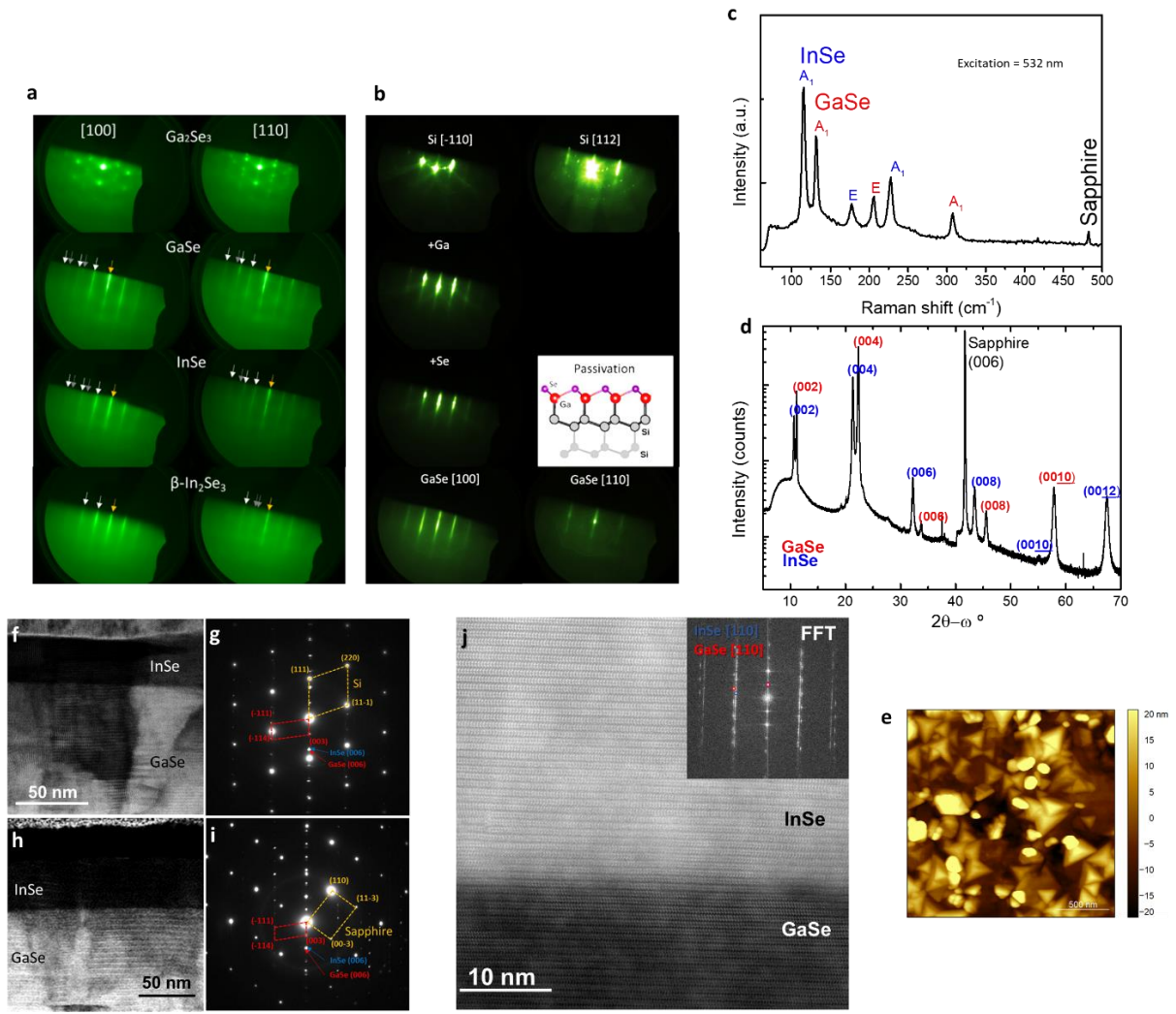

*Figure 2. a) In-situ reflection high-energy electron diffraction (RHEED) pattern of a) Ga$_x$Se$_y$ and In$_x$Se$_y$ in the different phases obtained on c-sapphire substrate, very similar patterns are obtained on Si(111) substrates without surface treatments. (b) "Twin-free" GaSe obtained on Si (111) substrates taking advantage of the Ga-Se passivation layer. c) Raman spectrum and d) X-ray diffractogram of the InSe/GaSe heterostructure (30 TL InSe on 60 TL GaSe) grown on c-sapphire, e) AFM image of the same InSe/GaSe heterostructure grown on Si (111) (2 × 2 μm$^2$). g) Bright-field (BF) STEM image of the InSe/GaSe heterostructure grown on Si (111). g) Selected area electron diffraction (SAED) pattern at the interface region between InSe/GaSe and silicon. h) BF-*

*STEM image of the InSe/GaSe heterostructure grown on c-sapphire (0001). i) SAED at the interface region between InSe/GaSe and c-sapphire. j) HAADF-STEM image of the interface region between the InSe and GaSe grown on Si (111). The inset in j) shows the Fast-Fourier Transformation (FFT) from the InSe/GaSe interface.*

We observe the coexistence of all the non-centrosymmetric TL ($D_{3h}$), named as ε-($P\bar{6}m2$), β-($P6_3/mmc$), γ-($R3m$), and δ-($P6_3mc$) polytypes and centrosymmetric TL ($D_{3d}$) polymorphs and their grain boundaries are present in samples grown on both substrates. It results in the observed contrast between regions of the same material in Bright Field (BF) of scanning transmission electron microscopy (STEM) (Figure 2e and g), and is expected to occur after twinning and stacking sequence changes, as happens in most of layered materials. The impact of these defects in future devices is still unknown, and attempts to reduce their formation in 2D materials[59,60], including GaSe[46], are under study.

Regardless of the rotation and twinning along the c-axis (around multiples of 30°)[56], the GaSe and InSe vdW layers stack nicely across the surface, as observed in the selected area electron diffraction (SAED) pattern (Figure 2g and i) and the high-magnification HAADF-STEM image (Figure 2j). Moreover, the interface between InSe and GaSe is well defined, over a vertical region of 1 to 3 TL. This result confirms that both materials grow epitaxially layer-by-layer on both substrates and with minimal In-Ga inter-diffusion, differently from a previous reports[48]. In fact, layer rotation, translational shear fault and local changes in the vdW interplanar distance are important mechanisms of stress release, which allows the quick change in the in-plane lattice parameters. Both materials are unstrained and exhibiting their natural lattice parameters (further discussed below) at their interface presented in the Figure 2i.

## 2.2 Quantum-wells and superlattices

Even though high-quality InSe on GaSe could be obtained with the conditions used above, the growth of GaSe on InSe and the growth of a GaSe/InSe/GaSe quantum well failed. In the trials to grow heterostructures using the same conditions used in the thick films, as soon as the Ga shutter is open, a sudden change in the RHEED is observed and the streaky diffraction pattern changes to a spotty pattern, typical of a rough surface and very similar to the pattern observed in the growth of $Ga_2Se_3$ (Figure 2a). A sample intended to have two 10 nm InSe quantum-wells separated by 20 nm GaSe barriers exhibited approximately a constant Indium signal along with the whole sample thickness (Figure 3b) when observed by X-ray photoelectron spectroscopy (XPS) depth profiling, indicating the formation of an $In_xGa_{1-x}Se$ alloy. In fact, in a separate experiment, where a 10 nm InSe layer is exposed to Ga (without Se) for sufficient time ( ~20 min), InSe transforms into (InGa)Se with very low In content (Raman in Figure 3c). Rather than diffusion, the exchange of Ga-In atoms during the growth is attributed to a well-known segregation (Figure 3a) MBE-effect as reported in InGaAs/GaAs QWs: where impinging Ga atoms replace In in the grown lattice, the replaced In stays on the surface, and wherein it is incorporated in the following layers.[61] The accumulation of In at the surface explains the change in the RHEED pattern that cannot result from simple diffusion.

As the In segregation is proportional to the growth temperature and Se flux[62,63], we modified the growth process for heterostructures, using substrate temperature of 380 °C during the growth of the InSe/GaSe interfaces. At lower temperatures the Se stick coefficient also change and the flux needs to be adjusted according in order to obtain the same phases. But at such low temperatures, the RHEED diffraction pattern loses part of its brightness, which points to an increase in the surface roughness. On the other hand, the RHEED remains streaky even after the growth of GaSe on InSe proceeds, indicating the continuity of flat vdW layer-by-layer growth and that the In segregation is significantly reduced. When the same stack with two 10 TL InSe quantum-wells was grown using 380 °C during the growth of the InSe QW and GaSe capping layers, the XPS depth profile shows In concentrated in the QW region (Figure 3b). Indeed, the characterization of this sample by STEM confirms the formation of a QW with well-defined interfaces within 1 to 2 TL and well-aligned layers along the plane. In the HAADF-STEM (Figure 3d) the contrast between InSe and GaSe matrix is clearly visible, and the composition is confirmed by energy-dispersive X-ray spectroscopy (EDX) in Figures 3e and 3f. In the XRD of 2x QW (Figure 3g), we note the appearance of week and broad satellites corresponding to its period (30 nm). The Raman spectrum (Figure 3h) exhibits some changes in comparison to the bulk/thick layers: a new mode at 125 cm$^{-1}$, the blue shift of ~3 cm$^{-1}$ in the GaSe $A_1$ mode, and in the InSe E mode, which could be attributed to the interaction between the distinct layers, similar to observations in $WS_2/MoS_2$ heterojunctions[64]. The shift or splitting of the E peak could also indicate the presence of $D_{3d}$ TL polymorphs[56] which is amply present in these STEM images (Figure 3d).

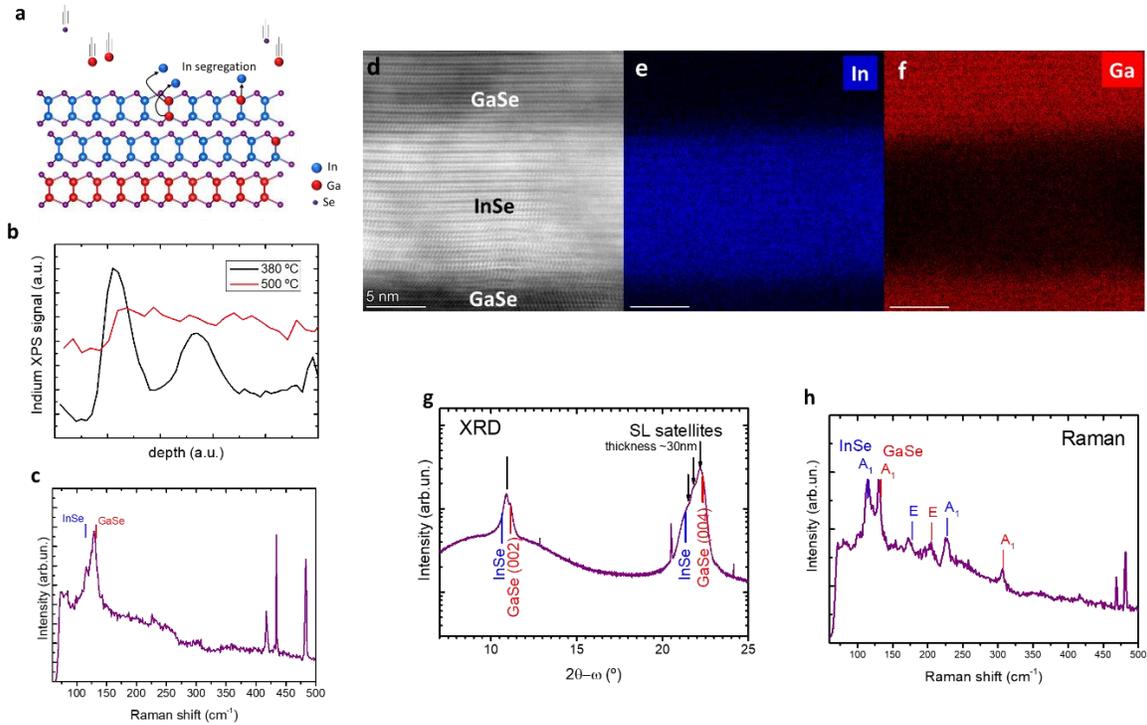

*Figure 3. a) Representation of In segregation during MBE growth. b) X-ray photoelectron spectroscopy (XPS) depth profile of two 20nm GaSe/10nm InSe QWs grown at 500 °C (red) and at 380 °C (black). c) Raman spectrum of 10 nm InSe exposed to Ga for 5 min without Se, which confirms the In segregation induced by Ga. d) HAADF-STEM image and the energy-dispersive X-ray spectroscopy (EDX) map of e) indium and f) gallium in the same image of the GaSe/InSe/GaSe QW grown on c-sapphire. g) X-ray diffraction and h) Raman spectrum of the 2x InSe/GaSe QWs grown on c-sapphire.*

With these modified conditions, we grown thick (>120 nm) superlattices containing alternating layers of approximately 2.5 and 4 TL (4 and 7 nm) of InSe and GaSe with an InSe/GaSe/InSe double quantum well in the center with an extra TLs (4/4/4 and 5/5/5 QW, respectively) as the one presented in the Figure 4f. The QW in the center is aimed to work as a preferential recombination center to obtain enhanced photoluminescence (PL) in these expected type-II heterostructures.

STEM shows sharp Ga and In distribution along the SL and QW (Figure 4a-e), with very well-defined interfaces and the expected layer thicknesses.

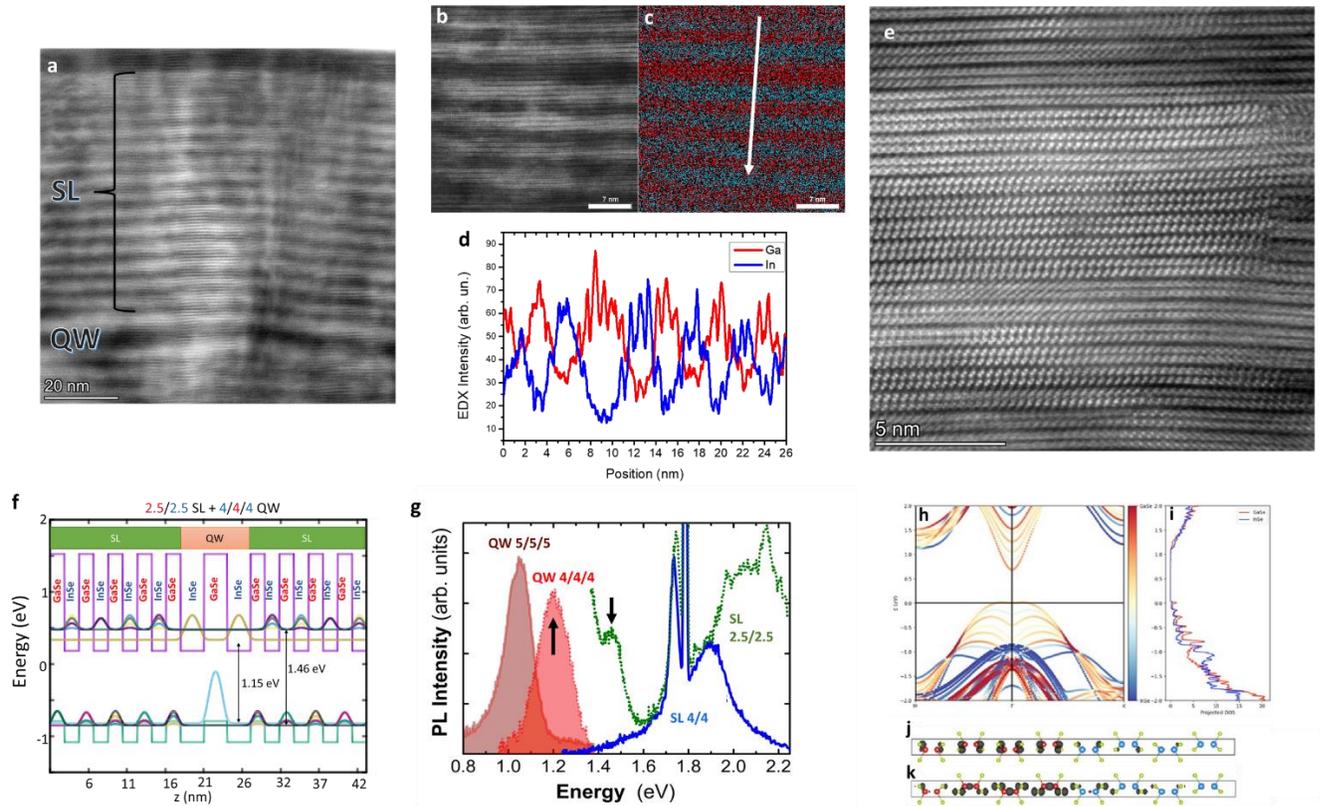

Figure 4. a) High-angle annular dark field (HAADF) STEM image of the 2.5/2.5 superlattice (SL) + 4/4/4 quantum well (QW) grown on c-sapphire. b) HAADF-STEM image and c) energy-dispersive X-ray spectroscopy (EDX) map of indium and gallium in the same image of the 2.5/2.5 superlattice (SL) + 4/4/4 quantum well (QW) grown on c-sapphire. d) Compositional profile of Ga and In obtained from the EDX along the white arrow in (c). e) Atomic resolution HAADF-STEM image 2.5/2.5 SL region. g) Simulated band profile (purple and green) along the growth direction, wavefunctions (modulus, colorful), and energies in the same sample using the function envelope and effective mass approximation (EF/EMA) g) Photoluminescence spectrum at 15 K of the 2.5/2.5 SL + 4/4/4 QW and 4/4 SL + 5/5/5 QW also grown on c-sapphire. (h) Electronic structure of the 4/4 TLs GaSe/InSe SL. The color map represents the weight corresponding to

*atomic orbitals from the 4 TLs of each material. (i) Projected density of states (pDOS) on GaSe (red line) and InSe (blue line). The electronic wave functions of the first conduction band state (j) and the first valence band state (k) are represented with isosurfaces in grey. Note that the wave functions are mostly localized around the selenium atoms (green).*

The most important observation in the PL spectra (Figure 4g) are the emission lines in the infrared region, peaked at 1.06 and 1.20 eV depending on the quantum well thickness. Particularly, we associate these PL lines to the optical interband transitions from the ground electron subband of the InSe/GaSe/InSe 5/5/5 and 4/4/4 TL QWs to the ground hole subband now confined in GaSe-QWs (5 and 4 TL thick). Given the lower value of the electron effective mass in the conduction band, most of the difference between those PL peak energies (140 meV) will be due to the difference in the electron confinement energy when the InSe QW narrows from 5 to 4 TLs according to our models. Both PL lines are weak, as compared to the signal measured in III-V semiconductor heterostructures, which is consistent with low absorption in InSe and GaSe materials[65] and the spatially indirect transitions due to the GaSe-InSe type II band alignment.

To obtain the direct (or quasi-direct) bandgap resulting from carrier confinement in the superlattice and QWs and compare them to the experimental data (Table 1) we used methods based on the function envelope and effective mass approximation[66] (EF/EMA). The EF/EMA model is widely used for Si-Ge and III-V semiconductors with high success and precision. The required values of the bandgap, effective mass ($m^*$), and valence-band offset (VBO) were collected from the limited literature available for PTMC heterostructures (Table 2). Strain is considered through deformation potential. Despite its simplicity, the method is accurate for wide quantum wells[67] (> 2 nm), showing a difference of only 4% to the experimental value or ab-initio calculations obtained from a 2/2 TL SL system.

Table 1. Experimental and calculated direct (Γ-Γ) bandgap for the bulk GaSe and InSe, its short-period superlattices, and embedded quantum wells.

|  | Experimental (eV) 15 K | EF/EMA (eV) 15 K |
|---|---|---|
| GaSe | (at RT) 2.0[a,b] | (at RT) 1.97[d] |
| InSe |  | (at RT) 1.26[d] |
| 2.5/2.5 TL SL | 1.45[a] | 1.46 |
| 4/4 TL SL | ~1.3[b] | 1.23 |
| 4/4/4 QW | 1.20[a] | 1.15 |
| 5/5/5 QW | 1.05[a] | 1.09 |
| 10/10 TL | 1.08[c] | (1.08)[e] |

a) Photoluminescence
b) Photoreflectance
c) Photocurrent
d) Reference [68] and [69]
e) Used for VBO determination

Table 2. Parameters used in the function envelope and effective mass approximation (EF/EMA). $m^*_e$ and $m^*_h$ are the electron and hole effective mass in the c-axis, respectively. α and β are the Varshini coefficients for temperature-dependent bandgap calculations[68]. $a_c$ and $a_v$ are the deformation potential for conduction and valence band in strained lattices.

|  | InSe | GaSe |
|---|---|---|
| $m^*_e$ | 0.14[70] $m_e$ | 0.17[71] $m_e$ |
| $m^*_h$ | 0.74[70] $m_e$ | 0.8[71] $m_e$ |
| Direct bandgap (0 K) | 1.35[68] eV | 2.10[69] eV |
| α (meV) | 0.475[68] | 0.66[69] |
| β (K) | 224[68] | 181[69] |
| $a_c$ (eV) | 8.9[72] | 5.3[73] |
| $a_v$ (eV) | 2.0[72] | 2.3[73] |
| Valence band offset (VBO) | -0.17 eV (See [74]) | reference |

Noteworthy, the comparison between InSe/GaSe/InSe 5/5/5 and 4/4/4 TL QWs using EF/EMA model (Figure S3 of the Supplementary Information) fully represent the observed in the PL measurement. The PL values similar to the value predicted by the EF/EMA model is strong evidence of carrier confinement in our heterostructures. Yet, in the same spectrum, it is possible to identify another peak that corresponds to the calculated 2.5/2.5 SL miniband energy in the EF/EMA model. Unfortunately, Photoreflectance (PR) measurements (Figure S1 at Supplementary Information) performed at room temperature exhibited poor signal-to-noise ratios, likely due to the inertness of vdW interfaces and the absence of significant band bending modulation upon excitation. Additionally, Fabry-Pérot interference turns it difficult to clearly define the peaks and their position. Nonetheless, there are indications of absorption in the energies pointed by the calculations and PL, between 1.0 and 1.5 eV, as well as, the absorption by the GaSe buffer and cap layer at 2.0 eV.

With the aim of better understand the band structure and valence band alignment of the heterostructure, we have performed ab-initio calculations using density functional theory (DFT) within the local-density approximation (LDA). Despite DFT typically underestimates the bandgap and is not always accurate for calculations of the conduction bands, DFT is a suitable tool to investigate ground state properties and therefore gives reliable values of the valence band alignment (or valence band offset) between at the GaSe/InSe interfaces. We have performed DFT-LDA simulations of a 4 TLs of InSe with 4 TLs of GaSe superlattices (with periodic boundary condition) and optimized the atomic distances and the in-plane lattice parameter. As measured by TEM, we find an intermediate lattice parameter, with the value of 3.88 Å (discussed in the next section).

Figures 4h-k show the electronic structure (h), density of states (i) and relevant wave functions of the GaSe/InSe superlattices. In order to observe the band alignment, we have represented the band structure using a color code that represents the weight of the atomic orbitals of each tetralayer. An electronic state totally localized at the GaSe (InSe) tetralayer will have a dark red (blue) color. As evidenced from the calculations, the electronic states near the bandgap are hybridized[75]

between the two layers with a slightly larger localization on the GaSe layer. For instance, the wavefunction of the top of the valence band state is clearly more localized on the GaSe layer (Figure 4j). On the contrary, the second valence band state is more localized on the InSe layer but still with a relevant hybridization. The conduction band state (Figure 4k) is expanded along both GaSe and InSe layers, because the electronic density is localized on the selenium atoms, which facilitates delocalization along the c-direction.

Our DFT-LDA calculations are also useful to estimate the valence band alignment. Thus, the valence band offset at the interface of two semiconductors is the energy difference between the top of the valence band of the two semiconductors resulting from the alignment of the Fermi level. This definition works well for quantum wells of a thickness of many atoms but it might be less precise when dealing with systems of few atoms, as in the present case. As established in our calculations, there is a non-abrupt change in the electron density and the wavefunctions are rather hybridized in the two layers[75]. Nevertheless, we can estimate the valence band offset (VBO) to 0.25 eV for a 4/4 TL SL and to 0.42 eV for a 3/3 TL SL, the difference between the state on top of the valence band and the next valence band state. This VBO value re-enforces the value found using the function envelope and effective mass approximation considering the strain and the PL measured energies.

### 2.3 Structure, Strain, and Sample homogeneity

The lattice parameters of the thick (bulk) film are extracted from localized diffraction (Figure 5 a and b), which is obtained through area-selected fast-Fourier transform (local-FFT) of the image of the interface containing both GaSe and InSe (Figure 2i). Measured interplanar distance of the peaks (003) and ($\bar{1}11$) were used to calculate a=b and c lattice parameters.

A similar method was used for the 2.5/2.5 SL. A profile of the (00n) and ($\bar{n}nn$) directions of the FFT (Figure 5e and f) points to a common in-plane lattice parameter a=b=3.88 Å (single peak), and distinct c=24.9 Å for GaSe and c=26.9 Å for InSe. The SL lattice parameters similar to the experimental ones were obtained in the previous DFT relaxations. Thereby, GaSe is in tensile strain and the InSe in compressive strain, assuming individually new interplanar distances and c-lattice parameters, larger than the relaxed bulk structure. Accordingly, a biaxial strain as high as 3% should be considered in precise theoretical models, the main effect is the increase of the valence band offset (VBO) and decrease of the type-II indirect bandgap. Nonetheless, as in the thick layers, the structure is indeed fully relaxed when the whole SL supercell is considered, due to strain compensation and relaxation.

In the heterostructure, the distinct layers exhibit in general the γ´-polymorph ($D_{3d}$ TL – See Figure 3d and Figure 4e), with a common in-plane lattice parameter. However in some locally-confined spots, others polytypes(ε- and β-polytype) and polymorphs (γ with $D_{3h}$ TL) are found. In fact, the strain likely causes the abundant presence of the rare γ´-polymorph[57]. The enthalpy of each GaSe polymorph, γ´ ($D_{3d}$ TL) and γ ($D_{3h}$ TL), was calculated using density functional theory for fully relaxed GaSe lattice, strained GaSe/InSe superlattices, and InSe surface lattice (Figure S2 in the Supplementary information). Similarly to observed on isolated TL[57], it was observed that the strain makes the γ´-polymorph slightly more stable and explains the abundance of the rare γ´ ($D_{3d}$ TL) form in the superlattice.

In the Raman spectrum (Figure 5g), the interfacial Raman mode (125 cm$^{-1}$) which is present in the previous single QWs sample (Figure 3h) is now enhanced. A new peak at 264 cm$^{-1}$ which corresponds to grey Se is visible, and again the shift or splitting of the E peak indicates the presence of $D_{3d}$ TL polymorphs[56].

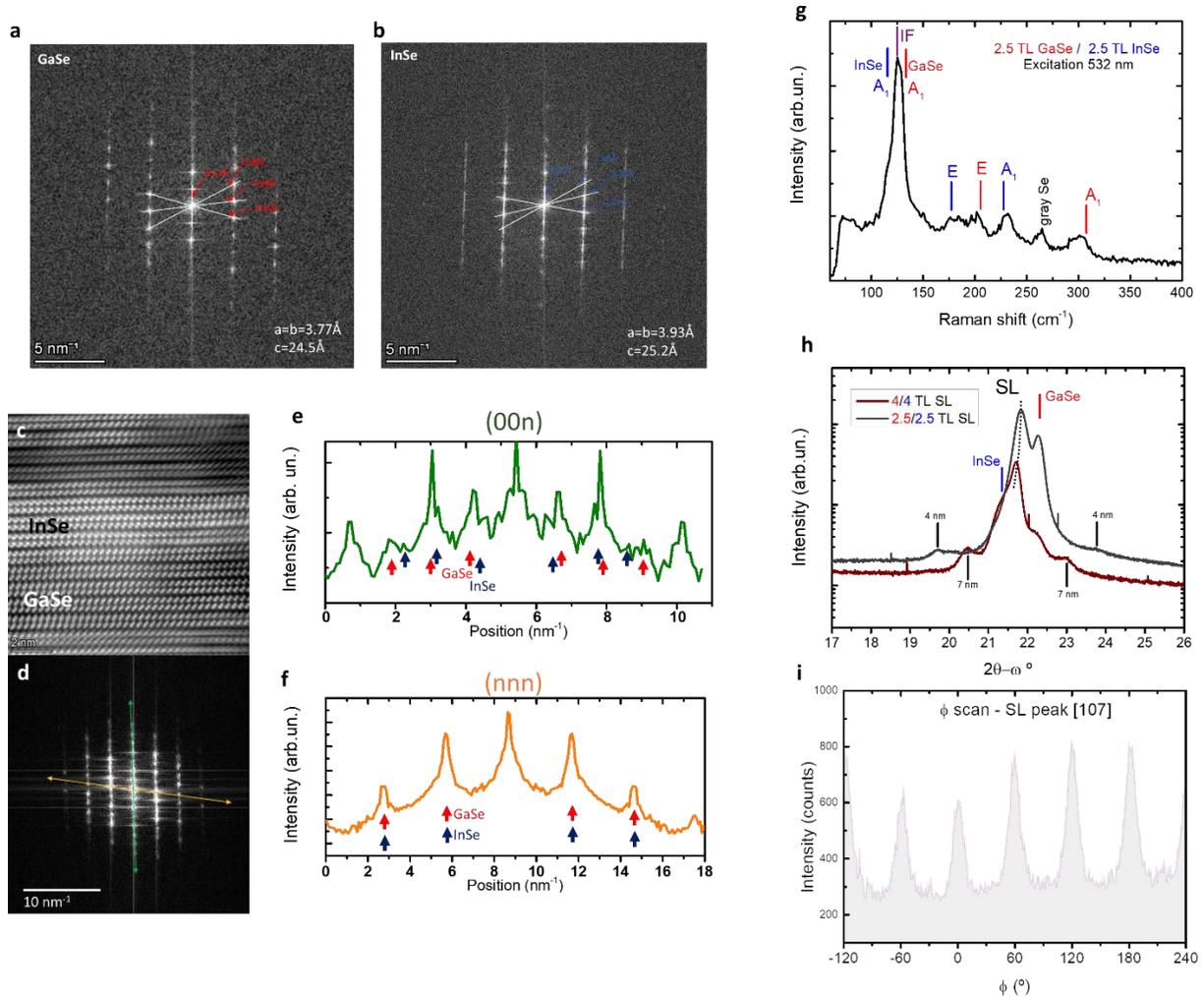

*Figure 5. Area-selected fast-Fourier transform (local-FFT) and measured interplanar distance (in colors) of GaSe (a) and InSe (b) in Figure 2j. c) HAADF-STEM image of the InSe/GaSe interface with the corresponding d) fast-Fourier transform (FFT), respectively. e) and f) The intensity profiles along the (00n) and ($\bar{n}nn$) planes are shown by the line arrows in (d). g) Raman spectrum of the 2.5/2.5 SL sample. h) X-ray diffraction of both samples which exhibit SL peaks with satellites*

*that closely correspond to the expected SL period and i) φ-scan of the peak (107) of the 2.5/2.5 TL SL sample.*

The MBE growth results in an overall film of uniform thickness with smooth surface morphology. The XRD (Figure 5h), which is taken in an area of 1 cm$^2$, present well defined SL satellite peaks, which assures the SL periodicity and interface quality over this extent. The BF-STEM images (Figure 2e and 2g) reveal the rotation of some of the GaSe and InSe domains, as well as the formation of grain boundaries between the domains, created during growth on both substrates. The selected area electron diffraction patterns (Figure 2f and 2h) obtained from the interfacial region between the InSe/GaSe films and the substrates show the following crystallographic relationships: InSe(001)/GaSe(001)/Si(111), GaSe(110)/Si(1-10), InSe(001)/GaSe(001)/sapphire(001) and GaSe(110)/sapphire(110). The same relationship is observed within 15 ° (full-width at half-maximum - FWHM) in the φ-scan of the peak (107) of the shortest period SL (Figure 5i). It states that notwithstanding the twin and polytypes grain boundaries, it is indeed a single crystal.

Despite the contrast variations in STEM (which is due to the FIB sample preparation, lamella thickness variation, and the grains slight off the zone-axis) it is noticeable that the homogeneity of the QW layer along the wafer is very satisfactory, as typical for the MBE method. Figure 6a shows the homogeneity of thick QW (10 nm) when seen at low magnification.

On the other hand, in the SL + QW sample, dark spots were visible on the surface, even in the optical microscope. The homogeneity of the photoluminescence signal from the QW (1.06 and 1.20 eV) means that it does not affect the core of the heterostructure and the variations are restricted to the surface. It was observed that in some of these spots, the PL signal attributed to grey selenium (1.74 eV) and its Raman (264 cm$^{-1}$) are stronger, pointing to an accumulation of Se on the surface.

In fact, selenium "bumps" were spotted on the surface in the STEM image (Figure 6b) and confirmed by local electron diffraction. The accumulation of Se at the surface is an issue for thick samples grown at low temperatures. It could be expected considering that the SL growth temperature is reduced, reducing also the re-evaporation of the excess of Se. It demonstrates the importance of precise control of the III/VI flux ratio. We believe that in the future the Se droplets on the surface can be removed with flash annealing inside the chamber during growth pauses or after growth, or just avoided controlling the cracker valve with higher precision.

It is also noticeable that defects (mostly polytypes/polymorphs grain boundaries, stacking faults, and dislocations) are more visible close to the substrate, in the 30-40 nm buffer layer, or just above it. It is related to the mismatch between the substrate and the GaSe lattice. Indeed, the first layers presented stretched in-plane lattice parameters (3.89 Å) to fit the sapphire lattice, which tends to relax creating the observed defects. However, these defects did not reach the QW and cause the complete suppression of PL signal as would be expected in some classical compound semiconductors. Nonetheless, there is room for improvements in the buffer layer that could reflect the PL and future device's performance. Considering that a fast relaxation was observed due to the vdW stack, a simple increase of the buffer layer thickness could be enough for initial improvements.

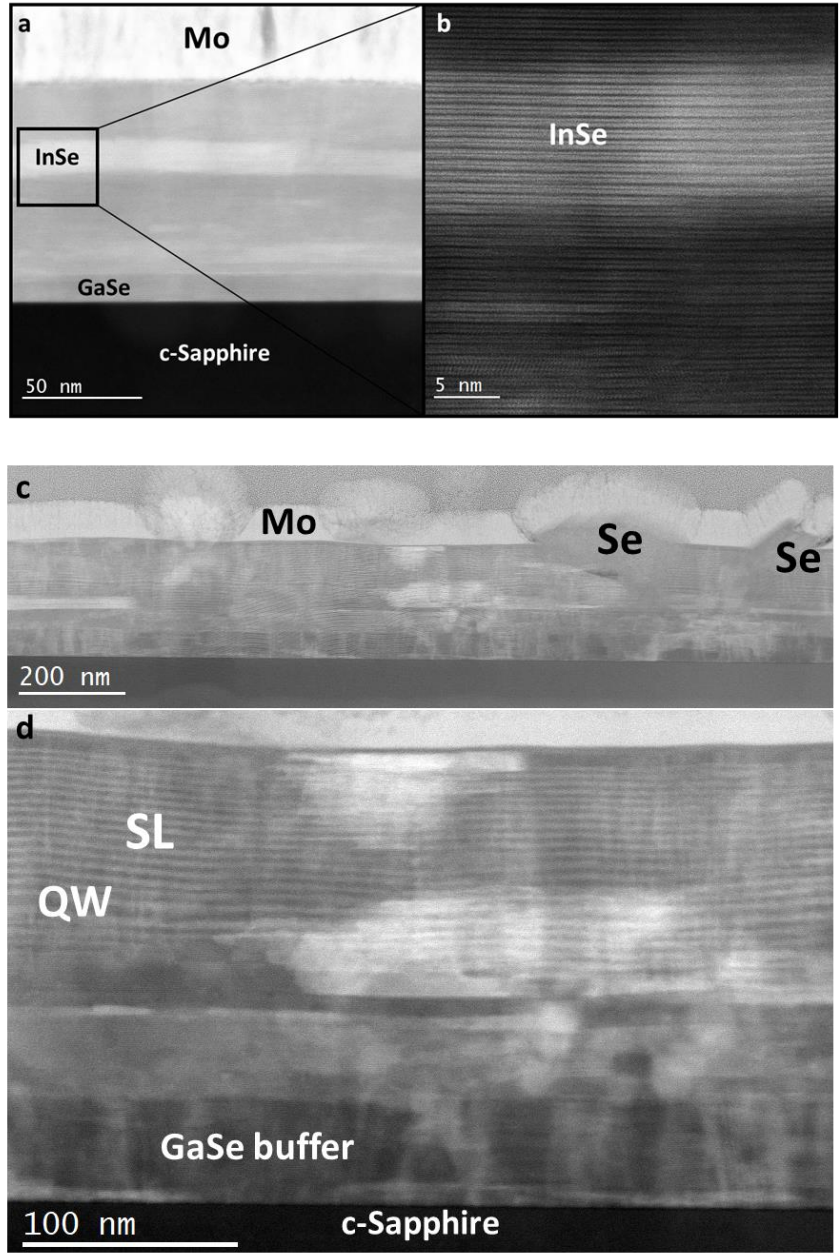

*Figure 6. a) Low-magnification and b) detail HAADF-STEM images of 10 nm InSe/20 nm GaSe QW. c) and d) HAADF-STEM images of the prepared lamella showing the c-Sapphire substrate, GaSe buffer layer, QW, SL, Mo, and the e-beam deposited Pt.*

**Photodiodes at wafer-scale**

To demonstrate the capabilities of our MBE growth process, proof-of-concept photodiodes were fabricated at wafer scale using standard microfabrication techniques. Our photodiodes are based on the vdW epitaxy of a 10 nm GaSe and 10 nm InSe heterojunction on boron doped $p^+$-Si (111) substrates and a transparent indium-tin oxide (ITO) top contact deposited by magnetron sputtering (see layer stack and expected band diagram[74,76–79] in Figure 7a). In order to protect the heterojunction from oxidation and preserve its properties, the photodiode area is defined by etching the stack perimeter by argon-ion milling until the substrate, and encapsulating with 200 nm of amorphous $Al_2O_3$. The processed wafer with several identical devices is presented in Figure 7b.

Unintentional doping is expected in GaSe and InSe grown by most methods, largely due to selenium vacancies, leading naturally to *p*-GaSe and *n*-InSe with majority carrier concentrations ranging from $10^{15}$ to $10^{17}$ cm$^{-3}$.[74,76,80] Consequently, the *p*-GaSe/*n*-InSe heterojunction is expected to show strong photo-response due to the built-in potential. Self-driven photodetectors that can detect light without any external voltage bias are important for low-power applications, including future internet of things, wearable and flexible electronics. 2D materials exhibit good optoelectronic properties; nonetheless, their extraordinary properties have not been fully exploited to realize high-performance, self-driven photodetectors due to the difficulties of producing such heterostructures at large scale.

While the *p*-Si/ITO junction exhibits an ohmic behavior[81], the fabricated device shows the existence of a rectifying junction (Figure 7c) that behaves as a photodiode with very-low threshold voltage (and $V_{oc}$) and responsivity peak of 60 mA/W at 800 nm without voltage bias (Figure 7d). This value is 2.5 times higher than the responsivity previously observed on similar devices based on flakes[74] and 1/10$^{th}$ of commercial p-i-n Si diodes, which usually have much thicker active regions, in the range of micrometers. Despite some can find reports of much higher responsivities[82], caution should be taken, since the interface defects can be the cause of the amplified responsivity usually seen in the 2D devices based on flakes and nanosheets (photogating[83]). Furthermore, the analysis of the photocurrent onset indicates that carrier photogeneration takes place at the type-II GaSe/InSe interface with the energy of 1.08 eV (1146 nm) which corresponds to a VBO of 170meV.

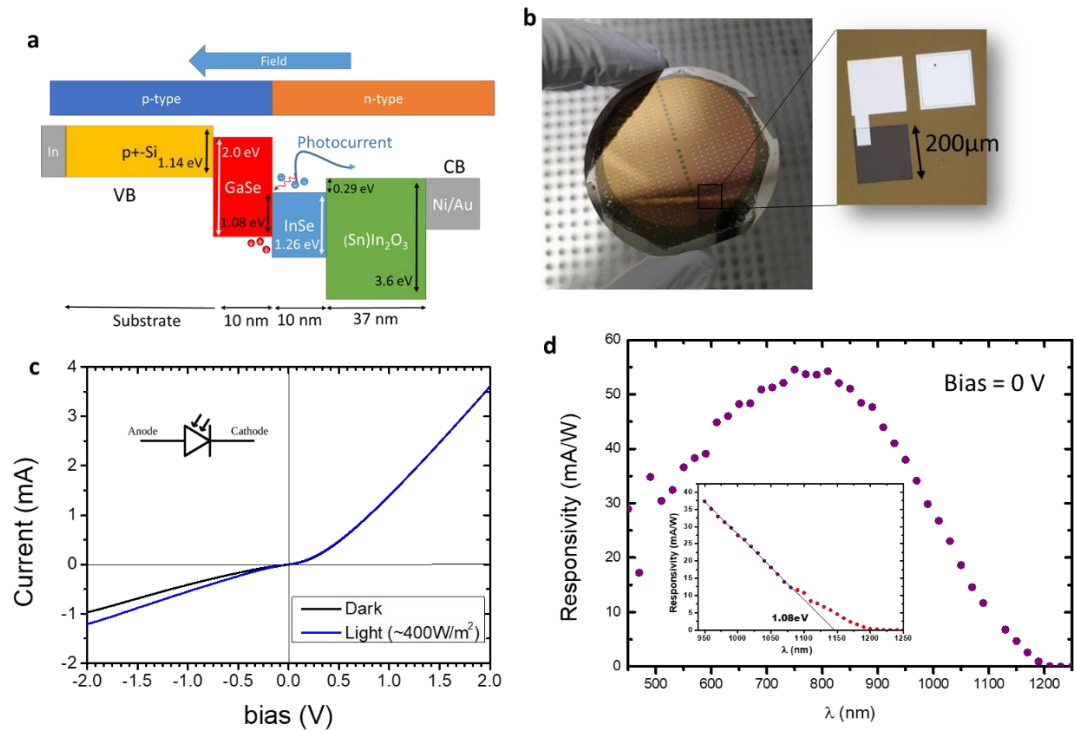

*Figure 7. a) Expected band diagram of the fabricated vdW photodetector. b) Photography of the 2-inch wafer after the fabrication process. c) IV curve and d) responsivity vs. wavelength curve of the device presented in (a) and (b).*

## 3. Conclusion

Van der Waals epitaxy of GaSe/InSe heterostructures containing quantum wells and superlattices has been demonstrated on Si and sapphire substrates using molecular beam epitaxy. Control of the crystalline phase and indium segregation was achieved by monitoring *in-situ* the surface properties by RHEED with simultaneous fine-tuning of the III/VI ratio and substrate temperature. The obtained interfaces are sharp and can be applied in the design of a large set of optoelectronic devices requiring quantum confinement, relaxing the requirement of lattice matching of mainstream Si-Ge, III-V or II-VI semiconductors.

These 2D materials not only cover a wide range of bandgap values, but more importantly, they form a novel class of materials exhibiting unparalleled properties such as high electron mobility, quantum Hall effect, or non-linear optical properties, and many other properties yet to be discovered. A significant improvement in the level of understanding about the doping[84–87], band structure, and defects in the GaSe/InSe system is still required before optoelectronic devices of similar quality as those based on Si or III-V technology can be fabricated. Nonetheless, our work presents a breakthrough as it demonstrates the controlled epitaxial growth of functional heterostructures and superlattices, thereby paving the way for a further rapid development. In fact, a disruptive displacement of Si technology by 2D materials is not to be expected, but instead, rather their coexistence and mutual benefits will prevail. The wafer-scale fabrication of a 2D self-driven photodetector on a Si substrate that can be included in the back-end of the line process flow is one proposal to realize it.

## 4. Experimental Section

**MBE growth:** The growth of GaSe and InSe was performed in an EVO-50 molecular beam epitaxy (MBE) system (Omicron Nanotechnology GmbH). Indium (6N) and gallium (6N) are evaporated from Knudsen cells at ~750 °C and ~850 °C, respectively. Selenium (5N) is evaporated from a valved cracker cell with reservoir maintained at 285 °C, while the flux is controlled by a valve with an adjustable aperture ranging from 0 to 8 mm. Before entering the growth chamber, large selenium molecules are cracked by the cracker stage kept at 900 °C. The stand-by base pressure of the MBE system is 2.6×10⁻¹⁰ mbar and during the growth, when the Se valve is open, the pressure increases to 10⁻⁸ to 10⁻⁷ mbar. All growth processes are monitored by reflection high-energy electron diffraction (RHEED, Staib Instruments), which was operated at 15 kV. Epi-ready single-side polished 2-inch c-sapphire (0001) and *p*-Si (111) substrates were used, with a specified and confirmed roughness of ~0.2 nm. Substrate temperatures are nominal (thermocouple at the heated surface) and close to the temperature of the Si substrate surface as measured by a two-color infrared pyrometer. The growth rate of GaSe and InSe is 1 TL/min (nm/min) calibrated ex-situ by X-ray reflectometry (XRR).

Each substrate was annealed in vacuum inside the growth chamber for 30 min at 950 °C just before the growth. Particularly in the Si substrate, the desorption of native $SiO_2$ and formation of the Si (111) 7×7 surface reconstruction was observed by RHEED. The samples presented in Figure 2 show GaSe layers grown on sapphire for 30 min. at 600 °C and 30 min. at 550 °C, and on silicon for 30 min. at 520 °C and 30 min. at 470 °C, with Se valve aperture of 1.15 mm. InSe thick layers were grown for 30 min. at 500 °C on sapphire and 450 °C on Si, with selenium valve aperture of 0.75 mm.

In the quantum-well samples, the first GaSe layer was grown on sapphire substrate for 10 min at 600 °C, 6 min at 550 °C with Se valve at 1.10 mm, plus 4 min while the temperature linearly ramps down to 380 °C, and the valve follows the temperature ramp from 1.1 mm to 0.65 mm. The InSe QWs were grown at the same temperature with Se valve at 0.7 mm. The following GaSe layers were grown at 380 °C and 0.65 mm. In the superlattice with embedded quantum well, the GaSe buffer layer is grown on c-sapphire for 23 min at 600 °C with Se at 1.15mm, 6 min at 550 °C with Se at 1.1 mm, and cooldown to 380 °C with Se at 0.7 mm, all the temperature changes occur during the growth at the rate of 60 °C/min and the Se valve changes follow linearly the temperature. Without pauses, the SL is grown at 380 °C alternating 15 (10) times 2 min (3 min) of GaSe and InSe in the case of the 2.5/2.5 TL (4/4 TL ) SL with Se at 0.7 mm for GaSe and 0.65 for InSe. The 4/4/4 (5/5/5) QWs are grown depositing each layer for 3 (4) min. The cap layer of 10 nm GaSe on top of the superlattice is grown at the same Se aperture yet at 380 °C.

**Material characterization:** Raman spectroscopy was measured at room temperature in a Witec alpha300 R confocal microscope, using a 50X objective lens, and a solid-sate 532 nm excitation laser. X-ray diffraction (XRD) measurements were performed in a PANalytical Xpert PRO MRD diffractometer with 5-axis cradle, standard Bragg-Brentano (BB) geometry, Cu anode X-ray tube operated at 45 kV accelerating voltage and 40 mA filament current to generate X-rays (Cu K-alpha). Soller and collimation 0.5" slits were used in the source side and a CCD detector (PiXcel) inline (1D) model with additional Soller slit. Atomic force microscopy (AFM) measurements were taken under ambient air conditions with a BRUKER Dimension Icon in tapping mode using PPP-NCH (Nanosensors$^{TM}$) cantilevers with a nominal tip radius of < 20 nm, force constant of 42 N/m, and ~ 265 kHz resonance frequency.

Cross-sectional TEM samples were prepared using standard lift-out procedure in a dual-beam FEI Helios NanoLab 450S focused ion beam with UHREM FEG-SEM. To protect the $In_2Se_3$ layers from oxidation, a Mo layer was deposited by sputtering immediately after MBE growth. Additionally, two protective Pt layers were deposited by electron beam and ion beam, respectively, to prevent Ga ion implantation and consequent damage during FIB preparation. Bulk milling and lamella thinning down to 100 nm was carried out using a 30 kV ion beam, while final thinning was performed using a 5 kV ion beam.The structural characterization was performed by STEM imaging using a FEI Titan Cubed Themis 60-300 kV double-corrected TEM/STEM at acceleration voltage 200kV and a convergence angle of 19 mrad. The high-angle annular dark field STEM

(HAADF STEM) images were acquired on the HAADF detector with 50.5 mrad inner and 200 mrad outer detection angles at a beam current of ~ 150 pA. Energy-dispersive X-ray (EDX) maps were acquired with a super-X EDX (FEI) system with simultaneous high-angle annular dark field (HAADF) image acquisition with 71.6 mrad inner to 200 mrad outer detection angles on the FEI HAADF STEM detector at a beam current ~ 500 pA to assure a satisfactory signal-to-noise ratio.

The XPS spectra were acquired with a hemispherical analyzer with pass energies 20 eV and 200 eV for high resolution and survey spectra, respectively. The XPS spectra were generated by an Al monochromated and a twin Al/Mg anode non-monochromated X-ray sources operated at 15 keV and power 200 W. The experiments were carried out in an ultra-high vacuum (UHV) system ESCALAB250Xi (Thermo Fisher Scientific). The base pressure in the system was below $5 \cdot 10^{-10}$ mbar. XPS spectra were peak-fitted using Avantage (Thermo Fisher Scientific) data processing software. For peak fitting Smart-type background subtraction was used. Quantification has been done using sensitivity factors provided by Avantage library. Elemental composition depth profiling was carried out by means of the monoatomic ion source (MAGCIS, Thermo Fisher Scientific). The sputtering rate established for $Ta_2O_5$ film was used to estimate the sputtering rate on the surfaces under study. The $Ar^+$ beam was raster-scanned over a 2 mm x 2 mm area. A dual beam charge neutralization technique with 0.5 eV Ar ions and 0.5 eV electrons was used for eliminated surface charging during the XPS measurements.

PL experiments at near infrared wavelengths (900 to 1700 nm) were carried out in backscattering geometry at low temperatures (15 K) by placing the samples in the cold finger of a closed-cycle cryostat (ARS model DE-202AE). The pumping was made at 532 nm by using a continuous wave DPSS green laser. Emitted light is coupled into a multimode optical fiber and detected by means of a NIRQuest512 spectrograph from Ocean Optics.

Room-temperature PR was carried out using the light beam of a quartz-tungsten-halogen lamp operated at 200 W. This probe light is passed through a monochromator (1/8 m Cornerstone-Newport) and focused with optical lenses on the sample. Light directly reflected with intensity $I_0(\lambda)R(\lambda)$ is focused on a solid-state detector, either Si or Peltier-cooled InGaAs. Two laser sources served as pump excitation, the lines at 325 nm of a 15 mW He-Cd laser (Oriel) and 632.8 nm of a 30 mW He-Ne (Melles-Griot). The pump beam is mechanically chopped at 777 Hz and superimposed onto the light spot of the probe on the sample, providing the periodic modulation. The current signal at the detector, containing the dc average signal $I_0(\lambda)R(\lambda)$ and the ac-modulated contribution $I_0(\lambda)\Delta R(\lambda)$ (where $\Delta R(\lambda)$ is the modified reflectance resulting from the modulated perturbation) is transformed into a dc-voltage and preamplified (Keithley). The complete signal feeds a lock-in amplifier (Stanford Instruments), which tracks the ac-signal at the chopping frequency. The relative change in reflectance is obtained thereof by normalizing the ac signal with respect to the dc component, with typical values in the range of $10^{-3}$ to $10^{-6}$.

**Computational method:** The function envelope and effective mass approximation (EF/EMA) model was calculated using the transfer-matrix method (TMM) in a customized open-source software previously tested on III-V and II-VI heterostructures. The *ab-initio* calculations were done using density perturbation theory (DFT) within the local-density approximation (LDA) and the norm-conserving pseudopotentials[88]. The DFT-LDA calculations has been done in 4GaSe/4InSe superlattices using an energy cutoff of 100 Ry and using a 9x9x1 k-grid.. In the suppl. Info the entropy was calculated have optimized lattice parameters and atomic positions until forces in each atom are smaller than 0.01 eV/Å. The basis-set cutoff energy is 160 Ry and the Brillouin zone integrated with 15x15x6 Γ-centered Monkhorst-Pack grid of k-points in the self-consistent calculations with convergence criteria of $1\times10^{-8}$ eV.

**Device fabrication:** The GaSe/InSe heterostructure was grown by MBE in a similar way as described in the Supplementary Information. The boron-doped $p^+$-Si(111) (resistivity = $10^{-1}$ $\Omega$.cm) substrate was annealed inside the MBE growth chamber for 30 min at 950 °C just before the growth. The first 10 nm GaSe layer was grown for 3 min at 520 °C, and 7 min at 470 °C with Se valve aperture of 1.15 mm. The 10 nm InSe layer was grown for 10 min at 450 °C with selenium valve aperture of 0.75 mm. The 37 nm ITO is deposited in a Kenosystec multi-target UHV sputtering system with 2" diameter magnetrons in confocal geometry, using 60 W RF plasma with pure argon atmosphere. The pressure in the chamber during the depositions was $3\times10^{-2}$ mbar. The wafer was transferred from one system to the other within few minutes of air exposure.

The device active area is then defined by argon ion milling (Nordiko 7500) after direct photolithography with AZ1505 photoresist using a 405 nm laser (DWL2000 - Heidelberg Instruments) and AZ400K developer. The etch depth is monitored by secondary-ion mass spectrometry (SIMS). Without removing the photoresist, 200 nm of $Al_2O_3$ was deposited in a Singulus Four-Target-Module (FTM) physical vapor deposition cluster tool, base pressure $6\times10^{-8}$ mbar, 1500W RF plasma with pure argon atmosphere and pressure of $5\times10^{-3}$ mbar during deposition. The photoresist and $Al_2O_3$ (lift-off) are removed from the active area surface with a $O_2$ plasma asher and acetone bath. The top 5 nm Ni/ 35 nm Au metallic contacts were deposited in the same multi-target UHV sputtering system and defined by lift-off using a similar photolithography procedure. The ohmic bottom contact is formed by welding indium metallic balls on the substrate surface with an iron tip at 380 °C.

**Device Characterization:** Photocurrent and IV curves were measured by a Keithley 6487 picoamperemeter in DC mode. For these experiments illumination was provided by a

QuantumDesign MLS-450-300 monochromator which light power density was measured using a calibrated Si PIN photodiode (Thorlabs DSD2).

ASSOCIATED CONTENT

The Supporting Information is available free of charge at …

PDF: Detailed description of samples growth, discussion about segregation, superlattice. Quantum-well characterization by XRD, Raman. Supplementary analysis of crystal structure and strain. Sample homogeneity. Band structure calculations.

DATA AVAILABILITY

The samples and data that support this paper and other findings of this study are available from the corresponding author upon reasonable request.

**Corresponding Author**

*Corresponding author: marcel.santos@usc.es

**Present Addresses**

†Centro Singular de Investigación en Química Biolóxica e Materiais Moleculares (CiQUS), Departamento de Química-Física, Universidade de Santiago de Compostela, Santiago de Compostela 15782, Spain

**Author Contributions**

M.S.C. coordinated the study, designed and executed the MBE growth, device processing, Raman spectroscopy, AFM, XRD, EF/EMA calculations and device characterization. A.M.-S. prepared the DFT calculations. A.B. performed the XPS depth profile, J.G., P.J.F. and K.E.H. executed the HR-STEM sample preparation and measurements. J.P.M.P and A.P.A performed the PL and D.F.M. the PR. S.S. supervised the study. All authors contributed to write the manuscript.

**Funding Sources**

Nanotechnology Based Functional Solutions (NORTE-01-0145-FEDER-000019)

Norte Portugal Regional Operational Programme (NORTE2020) - PORTUGAL 2020 European Regional Development Fund (ERDF).

Agencia Estatal de Investigación MCIN/ AEI/10.13039/501100011033/ - R&D project PID2020-113533RB-C31

RYC2018-024024-I, MINECO

Ministerio de Ciencia e Innovación, - Agencia Estatal de Investigación (AEI), project PID2020-112507GB-I00

**Notes**
The authors declare no competing financial interest.

ACKNOWLEDGMENT

This article has received support from the project Nanotechnology Based Functional Solutions (NORTE-01-0145-FEDER-000019), supported by Norte Portugal Regional Operational Programme (NORTE2020), under the PORTUGAL 2020 Partnership Agreement, through the European Regional Development Fund (ERDF). This work was supported by FCT, through IDMEC, under LAETA, project UIDB/50022/2020. The Spanish Ministry of Science and Innovation - Agencia Estatal de Investigación MCIN/ AEI/10.13039/501100011033/ is acknowledged for partial funding through the R&D project PID2020-113533RB-C31. A. M.-S.

thanks the Ramón y Cajal programme (grant RYC2018-024024-I, MINECO, Spain) and the Ministerio de Ciencia e Innovación, which is part of Agencia Estatal de Investigación (AEI), through the project PID2020-112507GB-I00 (Novel quantum states in heterostructures of 2D materials).

Supplementary Materials for

# Van Der Waals Heteroepitaxy of GaSe and InSe, Quantum Wells and Superlattices

*Marcel S. Claro†\*[1,2], Juan P. Martínez-Pastor[3], Alejandro Molina-Sánchez[3], Khalil El Hajraoui[1], Justyna Grzonka[1], Hamid Pashaei Adl[3], David Fuertes Marrón[4], Paulo J. Ferreira[1], Alex Bondarchuk[1], Sascha Sadewasser[1,2]*

[1] INL-International Iberian Nanotechnology Laboratory, Av. Mestre José Veiga s/n, 4715-330 Braga, Portugal.
[2] QuantaLab, 4715-330 Braga, Portugal
[3] ICMUV, Instituto de Ciencia de Materiales, Universidad de Valencia, P.O. Box 22085, 46071 Valencia, Spain
[4] Instituto de Energía Solar-ETSIT, Universidad Politécnica de Madrid, 28040, Madrid, Spain
\*Corresponding author: marcel.santos@usc.es

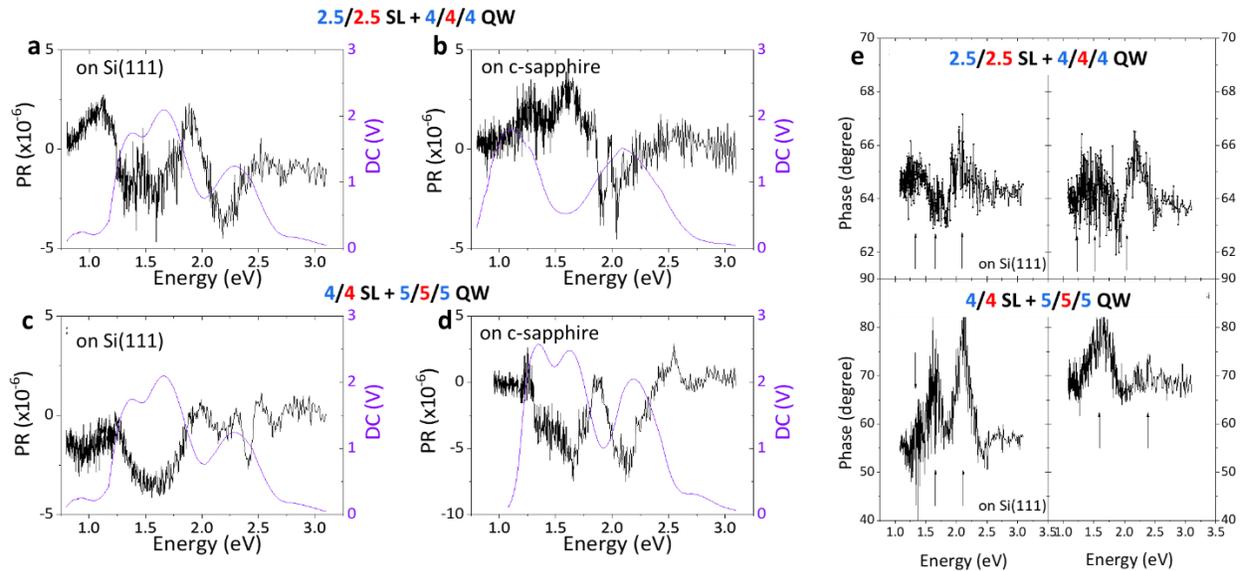

Figure S1. Photoreflectance spectra of a,b) 2.5/2.5 TL SL with 4/4/4 TL QW and the c,d) 4/4 TL SL with 5/5/5 TL QW on Si(111) (a and c) and c-sapphire (b and d) substrates. e) Shows the lock-in phase which helps to identify the multiple peaks.

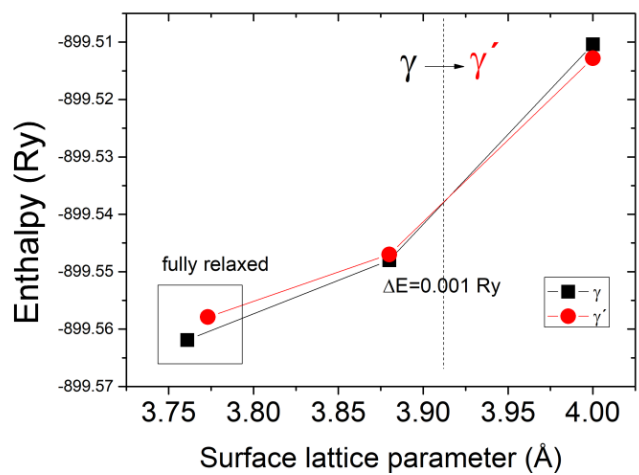

Figure S2: Ab-initio calculation of the enthalpy of fully and partially relaxed GaSe unit cell in the form of the γ´-polymorph ($D_{3d}$ TL) and the γ-polymorph ($D_{3h}$ TL).

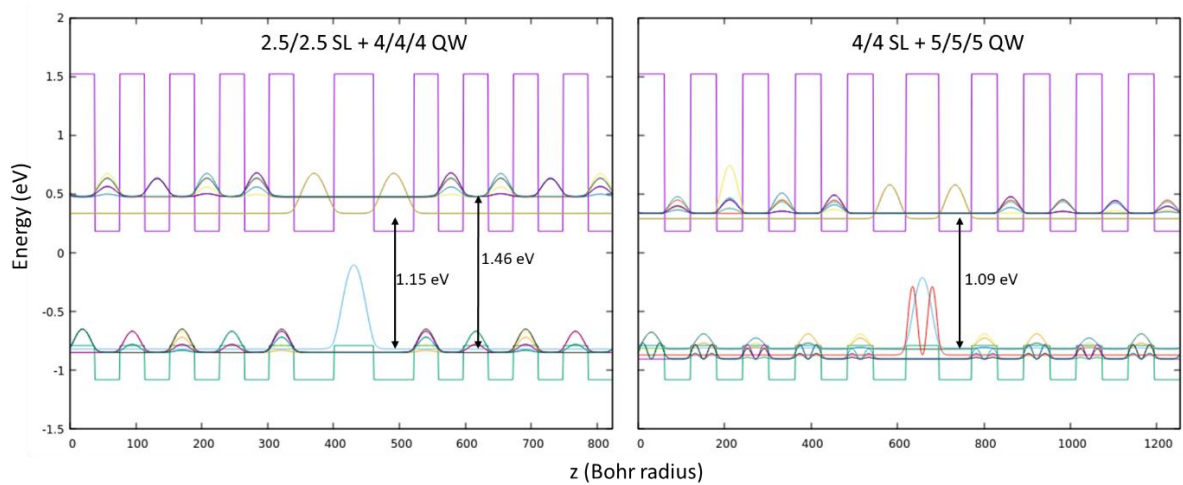

Figure S3. Simulated band profile (purple and green) along the growth direction, wavefunctions (modulus, colorful), and energies in the 2.5/2.5 SL + 4/4/4 QW (left) and 4/4 SL + 5/5/5 QW (right) samples which were calculated using the function envelope and effective mass approximation (EF/EMA).